\documentclass{elsart}
\usepackage{epsfig}

\journal{Nuclear Instruments and Methods A}

\begin{document}

\begin{frontmatter}

\title{
The multilevel trigger system of the DIRAC experiment}

\author[jinr]{L.Afanasyev\corauthref{cor1}},
\author[cern]{M.Gallas}, 
\author[basel]{D.Goldin},
\author[ihep]{A.Gorin},
\author[jinr]{V.Karpukhin},
\author[ioan]{P.Kokkas},
\author[jinr]{A.Kulikov},
\author[jinr]{K.Kuroda},
\author[ihep]{I.Manuilov},
\author[kyoto]{K.Okada},
\author[basel]{C.Schuetz},
\author[ihep]{A.Sidorov},
\author[basel]{M.Steinacher},
\author[kyoto]{F.Takeutchi},
\author[basel]{L.Tauscher},
\author[basel]{S.Vlachos},
\author[msu]{V.Yazkov}

\begin{flushleft}
\address[jinr]{Joint Institute for Nuclear Research, Dubna, Russia}
\address[cern]{on leave from CERN, Geneva, Switzerland}
\address[basel]{Basel University, Basel, Switzerland}
\address[ihep]{Institute for High Energy Physics, Protvino, Russia}
\address[ioan]{Ioannina University, Ioannina, Greece}
\address[kyoto]{Kyoto-Sangyo University, Kyoto, Japan}
\address[msu]{Nuclear Physics Institute, Moscow State University, Russia}
\end{flushleft}

\corauth[cor1]{Corresponding author:
E-mail: Leonid.Afanasev@cern.ch,\\
Phone: +7 09621 62539, Fax: +7 09621 66666\\
Mail address: Joint Institute for Nuclear Research, \\
Dubna, Moscow Region, 141980 Russia\\
Also at: CERN EP Division, CH-1211 Geneva 23, Switzerland}

\begin{abstract}
  The multilevel trigger system of the DIRAC experiment at CERN is
  presented. It includes a fast first level trigger as well as various
  trigger processors to select events with a pair of pions having a
  low relative momentum typical of the physical process under study.
  One of these processors employs the drift chamber data, another one
  is based on a neural network algorithm and the others use various
  hit-map detector correlations.  Two versions of the trigger system
  used at different stages of the experiment are described.  The
  complete system reduces the event rate by a factor of 1000, with
  efficiency $\geq$95\% of detecting the events in the relative
  momentum range of interest.
\end{abstract}

\end{frontmatter}

\section{Introduction} \label{sec:intr}

The DIRAC experiment \cite{proposal} at CERN measures the lifetime of
atoms consisting of $\pi^+$ and $\pi^-$ mesons ($A_{2\pi}$).  This
$A_{2\pi}$ lifetime is directly related to the difference $a_0 - a_2$
between the $s$-wave $\pi \pi$ scattering lengths with isotope spin
values 0 and 2.

This difference is calculated in the framework of the chiral
perturbation theory with high precision but has not been measured
experimentally with a corresponding accuracy yet. Theoretical
calculations predict the $A_{2\pi}$ lifetime to be close to $3 \cdot
10^{-15}$ s.

\begin{figure}[htbp]
\begin{center}
\epsfig{file=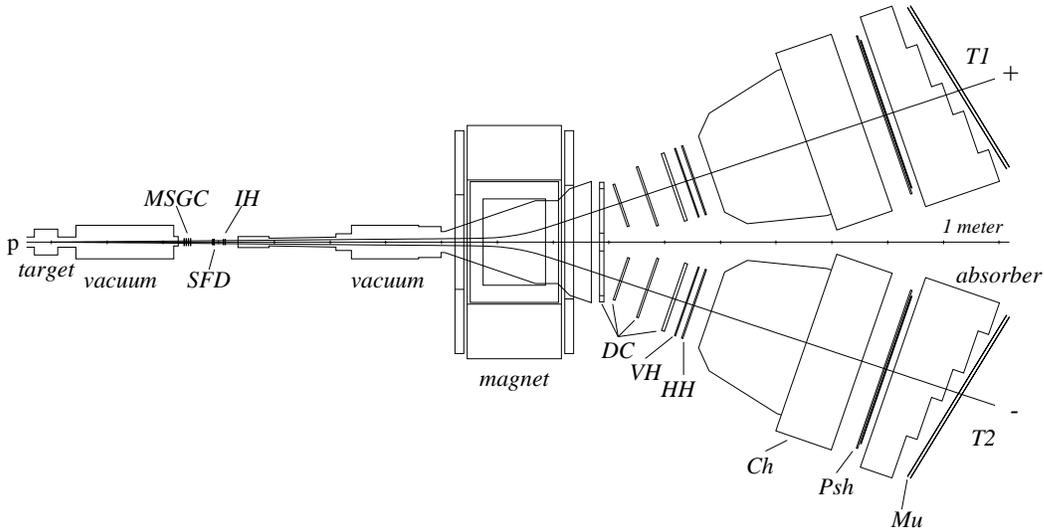,width=\textwidth}
\end{center}
\caption{Schematic top view of the DIRAC spectrometer. MSGC --
microstrip gas chambers, SFD -- scintillating fibre detectors,
IH -- ionization hodoscopes, DC -- drift chambers, VH,HH --
vertical and horizontal hodoscopes, Ch -- Cherenkov counters,
PSh -- preshower detectors, Mu -- muon detectors.}
\label{setup}
\end{figure}

The experiment is using the 24~GeV proton beam from the PS accelerator
at CERN. The DIRAC experimental setup (Fig.~\ref{setup}) is a magnetic
spectrometer with detectors arranged in one upstream and two
downstream arms. The spectrometer is placed in the secondary beam
produced by the primary protons hitting a fixed target. The upstream
part, along the secondary beam path before the dipole magnet, contains
microstrip gas chambers (MSGC), a scintillating fiber detector (SFD)
and a scintillation hodoscope (IH -- ionization hodoscope).  The
downstream arms include drift chambers (DC), scintillation hodoscopes
with vertically and horizontally oriented scintillators (VH and HH
respectively), gas Cherenkov counters (Ch), preshower detectors (PSh)
and finally muon counters (Mu) placed behind iron absorbers.

The $A_{2\pi}$ disintegrate into pairs of oppositely charged pions
with a very low relative momentum ($Q$), typically less than 3 MeV/$c$.
Such pion pairs have to be efficiently detected for the measurement of
the $A_{2\pi}$ lifetime.

Targets of different materials and thicknesses are used during data
taking. For each target the beam intensity is adjusted to obtain an
approximately constant counting rate.  The PS delivers the beam in
spills of duration 400--500~ms. With a typical proton beam intensity
of $10^{10}$ p/spill and a $94\mu$ thick Ni target, single rates in
the upstream detectors are about $3 \cdot 10^6$ counts/spill whereas
in the downstream detectors they are up to $8 \cdot 10^5$ and $6 \cdot
10^5$ counts/spill in the positive and negative arms respectively.

The trigger logic should provide a reduction of the event rate to a
level acceptable to the data acquisition system. Pion pairs are
produced in the target mainly in a free state with a wide distribution
over their relative momentum $Q$. The on-line event selection rejects
events with pion pairs having $Q_L>30\,$MeV/$c$ or $Q_x>3\,$MeV/$c$ keeping
at the same time high efficiency for detection of pairs with $Q$
components below these values.

\section{The trigger system}

A multilevel trigger is used in DIRAC. It comprises a simple and fast
first level trigger and several higher level trigger processors with
different selection criteria for various components of the relative
momentum of pion pairs.

Due to the requirements of the data analysis procedure the on-line
selection of only real pion pairs originating from a single
proton--target interaction is not enough. In addition, a large number
of uncorellated, accidental, pion pairs are also necessary. These
accidental pairs are used to reconstruct the relative momentum
distribution for free (non-atomic) pion pairs without Coulomb
interaction in the final state.  The measurement of this distribution
is indispensable for the correct calculation of the numbers of
produced and detected $A_{2\pi}$. Therefore, within a preselected
coincidence time window, the trigger system should apply very similar
selection criteria for real and accidental coincidences. The
statistical error of the $A_{2\pi}$ lifetime measurement depends on
the numbers of both real and accidental detected pairs. The optimal
ratio of real to accidental events at the present experimental
conditions is achieved with a $\pm 20$~ns coincidence time window
between the times measured in the left (VH1) and right (VH2) vertical
hodoscopes.

The trigger scheme was upgraded since the start of the experiment. In
this paper we present two trigger options: one was successfully used
during the first year of data taking and another one is being
currently used.

\begin{figure}[htbp]
\begin{center}
\epsfig{file=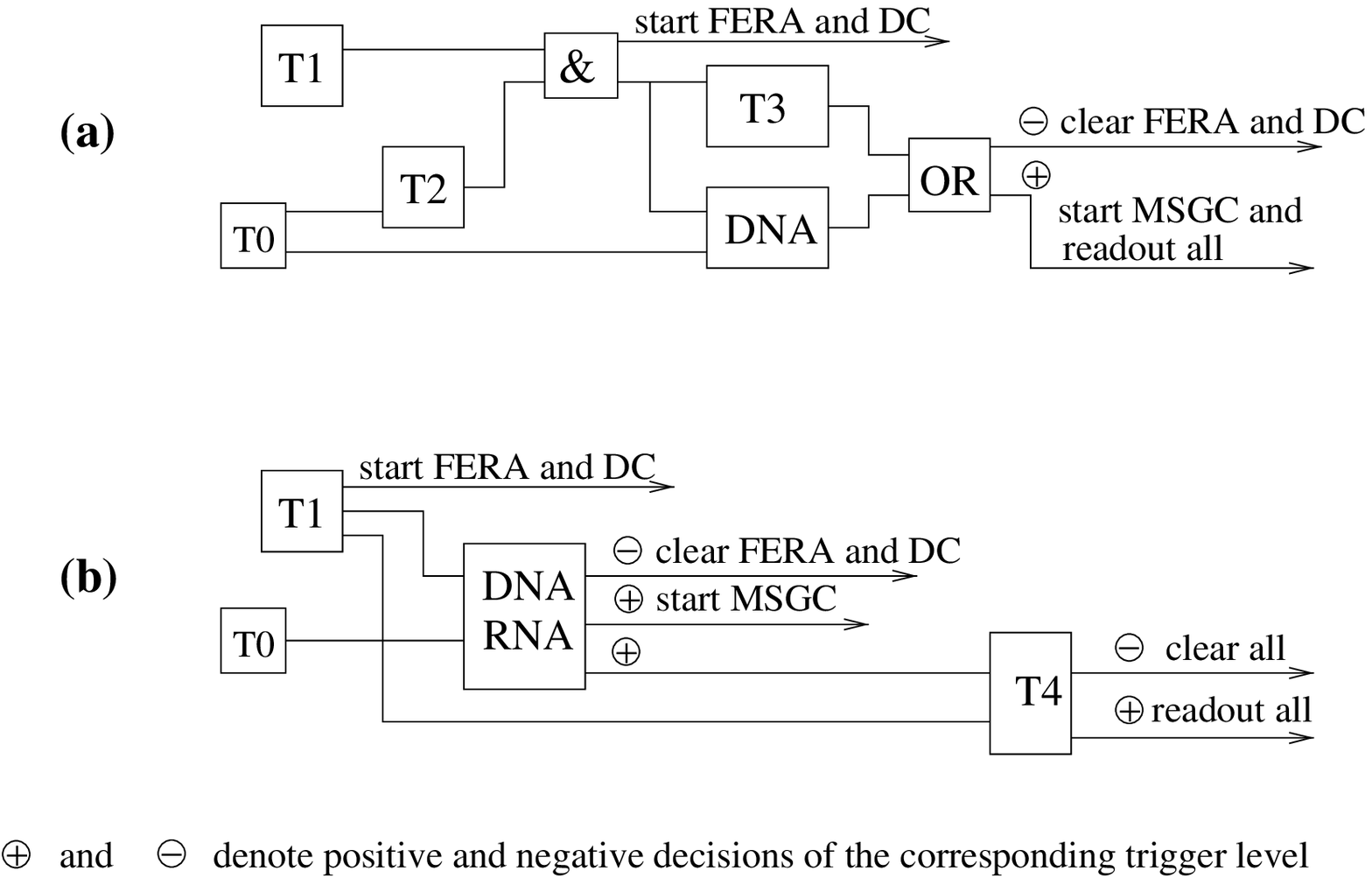,width=\textwidth}
\end{center}
\caption{General block diagram of the DIRAC multilevel trigger.
Variants {\bf a} and {\bf b} were used at the earlier and
later stages of the experiment, respectively.}
\label{trigab}
\end{figure}

The block diagram of the trigger structure is presented in
Fig.~\ref{trigab}.  In the diagram of Fig.~\ref{trigab}a the first level
trigger T1 in coincidence with the trigger T2 starts digitization of
the detector signals in the data acquisition (DAQ) modules (ADC, TDC
etc.). At the next level the OR of the trigger processors T3 and the
neural network trigger DNA (DIRAC Neural Atomic trigger) selects
further events by forcing additional constraints to the relative
momentum. The triggers T2 and DNA start with a fast pretrigger (T0).
A negative decision of T3 and DNA provokes clearing of all data
buffers and the trigger and data acquisition systems return to the
READY state.  A positive decision of the T3+DNA logic starts the
readout process.

In the latest option of the trigger system (Fig.~\ref{trigab}b) a
powerful drift chamber trigger processor T4 was added and the neural
trigger DNA was upgraded to a DNA/RNA version (Revised Neural Atomic
trigger) with improved performance. This made it possible to obtain
better trigger selectivity and higher efficiency.  Hence the T2 and T3
stages were not needed any more.

In addition to the mainstream trigger selection aimed at detecting
pionic atoms, there are also several special calibration triggers
which can be running in parallel with the main one or used separately
for dedicated measurements. They start the DAQ directly, bypassing the
selection of higher level triggers. When these triggers are used in
parallel with the main one appropriate prescaling factors are
implemented to maintain a reasonable total event rate.  Any trigger
stage higher than T1 can be disabled (set to a transparent mode).

\section{Trigger 1}

The first level trigger (T1) \cite{Trig1} fulfils the
following tasks:
\begin{enumerate}
\item Selects events with signals in both detector arms downstream the
  magnet;
\item Classifies the particle in each arm as a pion or an electron.
  Protons and kaons are equally included in the ``pion'' class. Muons
  could be identified with the use of the muon detector. However this
  detector is only used for off-line analysis and special dedicated
  measurements;
\item Arranges the coincidences between the two arms. The coincidence
  time window defines the ratio between the real and accidental events
  in the collected data;
\item Applies a coplanarity cut for pion pairs: a difference between
  the hit slab numbers in the horizontal hodoscopes in the two arms
  (HH1 and HH2, respectively) should be $\leq 2$. This criterion
  determines selection for the $y$ component of the pair relative
  momentum; 
\item Selects in parallel events from several physical processes
  needed for the setup calibration: e$^+$e$^-$ pairs, decays $\Lambda
  \rightarrow p + \pi^-$, $K^{\pm}$-decays to three charged pions,
  $\pi^+ \pi^-$ pairs without the coplanarity cut.
\end{enumerate}

All T1 modules are ECL line programmable multichannel CAMAC units.
Most of them are commercial modules except the dedicated coplanarity
processor developed at JINR.  To reduce the time jitter of the
trigger, meantimer units removing the dependence of timing on hit
coordinate are used in all VH and HH channels.

The pion ($\pi1,\pi2$) and electron (e1,e2) signatures in each of the
downstream arms are defined as a coincidence between different
detectors:
$$
\mathrm{VH1(2)} \cdot \mathrm{HH1(2)}  
\cdot \overline{\mathrm{Ch1(2)}} 
\cdot\mathrm{PSh1(2)} = \pi1(\pi2)
$$
$$
\mathrm{VH1(2)} \cdot \mathrm{HH1(2)}  
\cdot \mathrm{Ch1(2)} \cdot\mathrm{PSh1(2)} = 
\mathrm{e1(e2)}. 
$$
Here VH1, VH2 etc. denote the logical OR of all signals from the
corresponding detector.  The signatures from both arms are combined to
produce the final first level trigger. The timing of ``$\pi$'' and ``e''
signals is defined by the vertical hodoscope of the corresponding arm,
as the VH has the best time resolution among all detectors. At the
trigger level VH slabs are aligned in time with an accuracy of 1~ns.
Further off-line time corrections provide 175~ps resolution for the
time difference between the signals in the two arms.

The definitions of the subtriggers are the following.
\begin{itemize}
\item The pion pair ``atomic'' trigger: $A_{2\pi} = \pi1 \cdot \pi2
  \cdot \mathrm{Copl}$ \footnote{At the early stage of the
    experiment the signal of the ionization hodoscope IH participated
    in coincidences in this and all other T1 subtrigger modes. Later
    it was excluded due to its high occupancy.}, where ``Copl'' is the
  positive decision of the coplanarity selection processor.  The
  coplanarity processor (15~ns decision time) reduces the trigger rate
  by a factor of 2.
\item The electron pair trigger: ${\mathrm{e}^+ \mathrm{e}^- =
  \mathrm{e}1 \cdot \mathrm{e}2}$.
\item The pion pair trigger (no coplanarity selection):
$\pi^+ \pi^- = \pi1 \cdot \pi2$.
\item The $\Lambda$-decay trigger $\Lambda \rightarrow p + \pi^-$:
$$
\Lambda = (\mathrm{VH}1[17] \cdot \mathrm{HH}1 \cdot 
\overline{\mathrm{Ch1}} \cdot \mathrm{PSh1})
(\mathrm{VH}2[1 \div 16] \cdot \mathrm{HH}2 \cdot 
\overline{\mathrm{Ch2}} \cdot \mathrm{PSh2}).
$$
As a matter of fact, this is the same definition as that for the
$\pi^+ \pi^-$ trigger but here only vertical hodoscope slab 17 in VH1
and slabs $1 \div 16$ in VH2 out of the total 18 slabs of each arm are
used. This reflects the kinematics of $\Lambda$-decay where the proton
(in arm 1) and the $\pi^-$ (in arm 2) hit very different VH slab
ranges.
\item The $K$-decay trigger ($K^+ \rightarrow \pi^+ \pi^+ \pi^-$,
$K^- \rightarrow \pi^- \pi^- \pi^+$):
$$
\mathrm{K} = \pi1 \cdot \pi2 \cdot Maj[\mathrm{VH} \geq 
3] \cdot
Maj[\mathrm{HH} \geq 3] \cdot Maj[\mathrm{VH} < 5], 
$$
where $Maj$ denotes the majority logic applied to the number of
hits in the slabs of both VH or both HH hodoscopes. Thus, from the
pions detected in two arms only the events with at least 3 particles
in the downstream detectors are selected and at the same time too
complicated events with a high multiplicity $\geq 5$ are rejected.
Simultaneous majority selection in the vertical VH and horizontal HH
hodoscopes helps to suppress edge-crossing of the adjacent hodoscope
slabs by single particles which could imitate the passage of two
independent particles in one hodoscope.
\end{itemize}

For the last two subtriggers ($\Lambda$ and $K$) the coincidence time
window is reduced by a factor of 2.5 as there is no need to collect
accidental events for these categories. It is evident that the
$\Lambda$- and $K$-triggers do not provide detection of clean
$\Lambda$ and $K$ events but they enhance their proportions in the
final data sample.

\begin{figure}[htbp]
\begin{center}
\epsfig{file=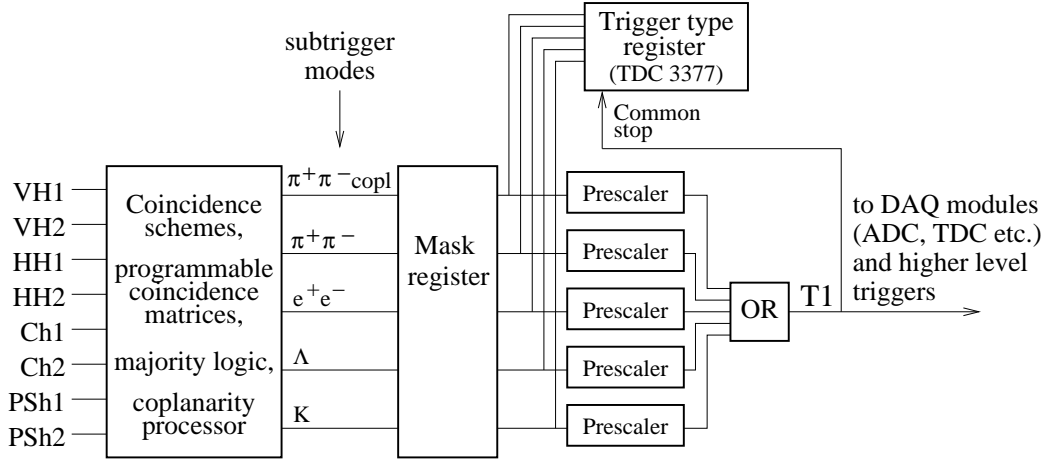,width=\textwidth}
\end{center}
\caption{Combination of different subtrigger modes in the 1st level trigger.}
\label{fig3}
\end{figure}

The block diagram of the combination of various subtriggers is shown
in Fig.~\ref{fig3}.  All signals pass through the mask register and
after prescaling are combined with an OR function.  Any subtrigger can
be enabled or disabled by proper programming of the mask register.
Timing of all subtriggers is the same.  Independent prescaling of each
subtrigger channel allows one to adjust the relative rates and to keep
the ratio of the main and calibration trigger rates at an optimum
level.  A specific trigger mark is recorded for every event, so the
data can be sorted by trigger type at the off-line analysis and
on-line monitoring.

The final first level trigger signal T1, as shown in Figs.
\ref{trigab}a and \ref{trigab}b, initiates operation of the DAQ
modules (gating of ADC, starting of TDC etc.) either directly or after
a coincidence with the T2 decision.  It also triggers higher level
processors.  Depending on the decisions of the trigger processors the
event data will either be converted and moved to the data collection
memories or discarded.  During this period generation of a new T1
signal is inhibited.

\section{Pretrigger T0}

Some trigger processors, especially T2 and DNA/RNA, need a fast
initial signal to start evaluating an event. An early fast pretrigger
(T0) is provided for that purpose. It is a simple coincidence of
signals from at least one slab in each of the VH1, VH2, PSh1 and PSh2
detectors, the coincidence time window being the same as for the T1
trigger ($\pm 20\,$ns).  To obtain a T0 signal as early as possible the
VH signals are taken directly from the discriminator outputs of the
upper PMs, i.e. before the meantimers. As a consequence, T0 has a 3~ns
jitter which disappears in the final trigger due to further
coincidence of various decisions with T1.

\section{Trigger 2}

Trigger 2 (its general ideas are considered, in particular, in
\cite{trig2}) uses the upstream SFD and IH detectors to select pairs
with small distances along the $x$ direction ($\Delta x$).  This leads
to rejection of events with a high component of the relative momentum
along $x$ ($Q_x$).  Typical relative momenta of pions from the
$A_{2\pi}$ breakup correspond to particle distances $\Delta x \leq
9$\,mm in the upstream detectors.

\begin{figure}[htbp]
\begin{center}
\epsfig{file=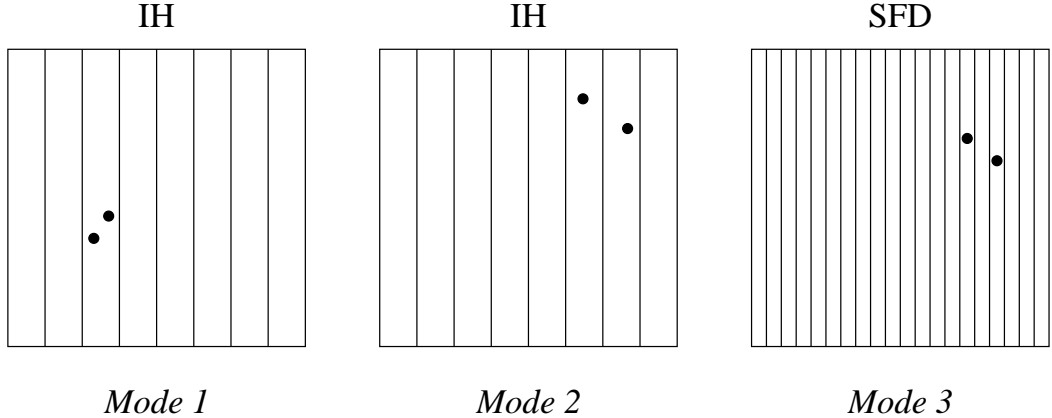,width=\textwidth}
\end{center}
\caption{Different T2 trigger modes. \emph{Mode\,1}: $\Delta x 
\leq 6$\,mm in one plane. In two staggered planes $\Delta x
\leq 3$\,mm. \emph{Mode\,2}: $\Delta x \leq 12$\,mm. \emph{Mode\,3}:
$\Delta x \leq 9$\,mm.}
\label{fig4}
\end{figure}

The T2 trigger includes three independent modes combined by an OR
function (Fig.~\ref{fig4}): in \emph{Mode\,1}\, and \emph{Mode\,2}\, the
data from two IH planes (A and B) are used and in \emph{Mode\,3}\, the
information from SFD is evaluated.  Only the X-planes of each detector
are used.

\emph{Mode\,1} selects events with two charged particles hitting the
same slab in the plane X-A of IH and the corresponding element in the
plane X-B. Selection of double hits is based on charge discrimination
with a threshold corresponding to detection of double ionization. An IH
scintillator slab is 6 mm wide, therefore a double ionization signal
in a plane arises if $\Delta x \leq 6$ mm.  However the plane X-B is
shifted with respect to X-A by half-width of a slab, hence only
particle pairs with $\Delta x \leq 3$\,mm are accepted by \emph{Mode\,1}.

To ensure the same selection efficiency for two uncorrelated particles
as for pions originating from a $A_{2\pi}$ breakup, dedicated
integrators have been designed and implemented in all IH channels of
the X-planes.  The integration is started by T0. The integrator
accumulates the input charge during a preset time interval and then
compares it with the double ionization threshold.

The outputs of the double ionization threshold discriminators are
connected to a trigger logic circuit (TLC) which requires a
coincidence between the double ionization signal from a slab in the plane
X-A with a signal from the matching slab in X-B. As the slabs in the
planes X-A and X-B are staggered, the coincidence in TLC is performed
for two possible matching combinations.

In \emph{Mode\,2} hits (of any amplitude over the single ionization
level) in adjacent slabs of planes A or B of IH are required. This
mode accepts the events with $\Delta x \leq 12$ mm.

\emph{Mode\,3} provides independent selection of events with small
$\Delta x$ using the SFD \cite{SFDtrig}. Signals from the fiber
columns (the fiber diameter is 0.5\,mm) of the SFD X-plane after
peak-sensitive discriminators \cite{peaksens} are coupled in pairs and
come to a SFD trigger logic circuit (TLC). TLC tests if hits with
distance $x \leq$9\,mm are present in the SFD hit pattern.  This value
of the $\Delta x$ interval has been chosen as it coincides exactly
with the range of distances for pions from $A_{2\pi}$ dissociation
(the distance range in TLC can be varied from 1 to 15\,mm). However,
the SFD trigger efficiency drops at $\Delta x < 1$\,mm as the tracks
cannot be resolved when two particles hit the same fiber column or a
pair of adjacent coupled columns.

Decisions of all 3 modes are combined with an OR function in order to
increase the trigger efficiency: \emph{Mode\,1} recuperates the events
rejected by \emph{Mode\,3} due to very small $\Delta x$, \emph{Mode\,1}
and \emph{Mode\,2} compensate inefficiency of the SFD detector itself
(which is not negligible) while \emph{Mode\,3} recovers the events lost
by \emph{Mode\,1} and \emph{Mode\,2} due to presence of small gaps
between the IH scintillators.

The resulting global T2 signal is synchronized with a delayed T0
signal; this decreases the time jitter of T2 decisions down to the
3~ns jitter of T0. In further coincidence with T1 the timing of T1$\cdot$T2
is defined by the T1 signal.

The selection efficiency for low $Q$ events of T2 is around 90--95\%,
and its rejection power is about 1.3--1.4. Both numbers vary slightly
according to the beam conditions (beam time structure and background
level).

\section{Trigger 3}

Further rejection of high $Q$ events is provided by the T3 processor.
Its implementation initially considered in \cite{T3note} is presented
in detail in \cite{Trig3}.  T3 makes a fast analysis of hit patterns
in the vertical hodoscopes VH1, VH2 and the ionization hodoscope IH.

\begin{figure}[htbp]
\begin{center}
\epsfig{file=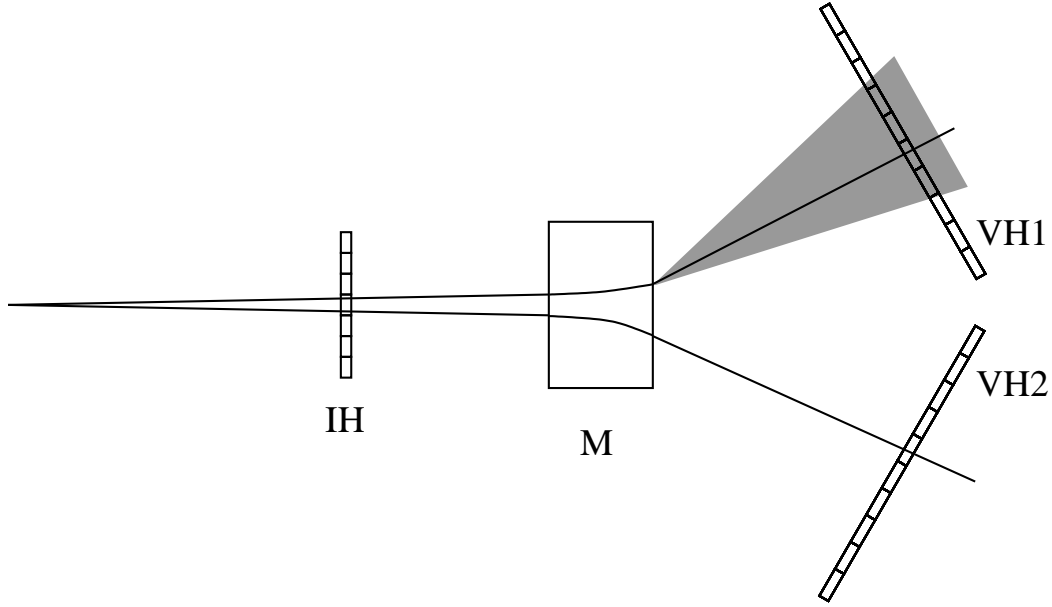,width=\textwidth}
\end{center}
\caption{Logic of the T3 selection.}
\label{t3logic}
\end{figure}

For pairs with low longitudinal component of the relative momentum
($Q_L$) there is a correlation between the hits in VH1 and VH2 (see
Fig.~\ref{t3logic}): for any slab hit in VH1 there is a limited group
of slabs in VH2 where the corresponding hit could be found. The
algorithm becomes much more selective if the information about the
slabs hit in the ionization hodoscope IH is additionally used to define
the particle's X-coordinates before the magnet (obtained from IH hit
slab numbers) and consequently indicate the appropriate VH1--VH2
correlation map.  T3 applies a $Q_L \leq 30\,$MeV/$c$ cut in
its current configuration.

The input data for T3 are two 18-bit patterns from the meantimers of
the VH1 and VH2 hodoscopes and a 16-bit pattern from IH.  The IH
pattern used is produced by the logic of the double ionization event
selection (as in T2). It provides the X-coordinate region(s) where
double ionization in one slab or two hits in adjacent slabs are
detected (similar to \emph{Mode\,1}+\emph{Mode\,2} of T2 but requesting
double ionization in any of the IH X-planes).

The T3 logic is based on LeCroy Universal Logic Module 2366. The
Xilinx FPGA (Field Programmable Gate Array technology) chip of this
module has been programmed with the corresponding trigger algorithm.
The implemented correlation maps of the VH1, VH2 and IH signals were
obtained with Monte-Carlo simulation of the DIRAC setup and further
tested with real experimental data.

The module configuration is stored in a binary file. Before data
taking it is loaded into the Xilinx chip via the CAMAC bus.  Files
with different configurations may be downloaded to adapt the T3
performance to specific needs.  The input signals to T3 are latched by
T1 (or T1$\cdot$T2) and 120~ns later the decision of T3 about the
event is delivered.

There is also a special ``transparent'' T3 operation mode.  A signal in
the dedicated ``transparent'' input forces T3 to accept the event. This
mode is used for calibration triggers ($e^+e^-$, $\Lambda$ etc.)  when
there should be no event suppression by T3: these event flags come
into the ``transparent'' input and hence prevent the event from being
probably rejected.

The T3 rejection power for typical experimental conditions is around
2.0 with respect to T1. The T3 efficiency for pairs with $Q_x \leq
3\,$MeV/$c$, $Q_y \leq 3$\,MeV/$c$ and $Q_L \leq 30\,$MeV/$c$ is 97\%.

\section{DNA and RNA neural network triggers}

The DNA (DIRAC Neural Atomic) trigger \cite{DNA} is a processing
system based on a neural network algorithm. The DNA hardware is based
on the custom-built version of the neural trigger used initially in
the CPLEAR experiment \cite{neural}. The flexibility of the
implemented algorithm allowed the system to be incorporated in the
DIRAC trigger system too.

\begin{figure}[bp]
\begin{center}
\epsfig{file=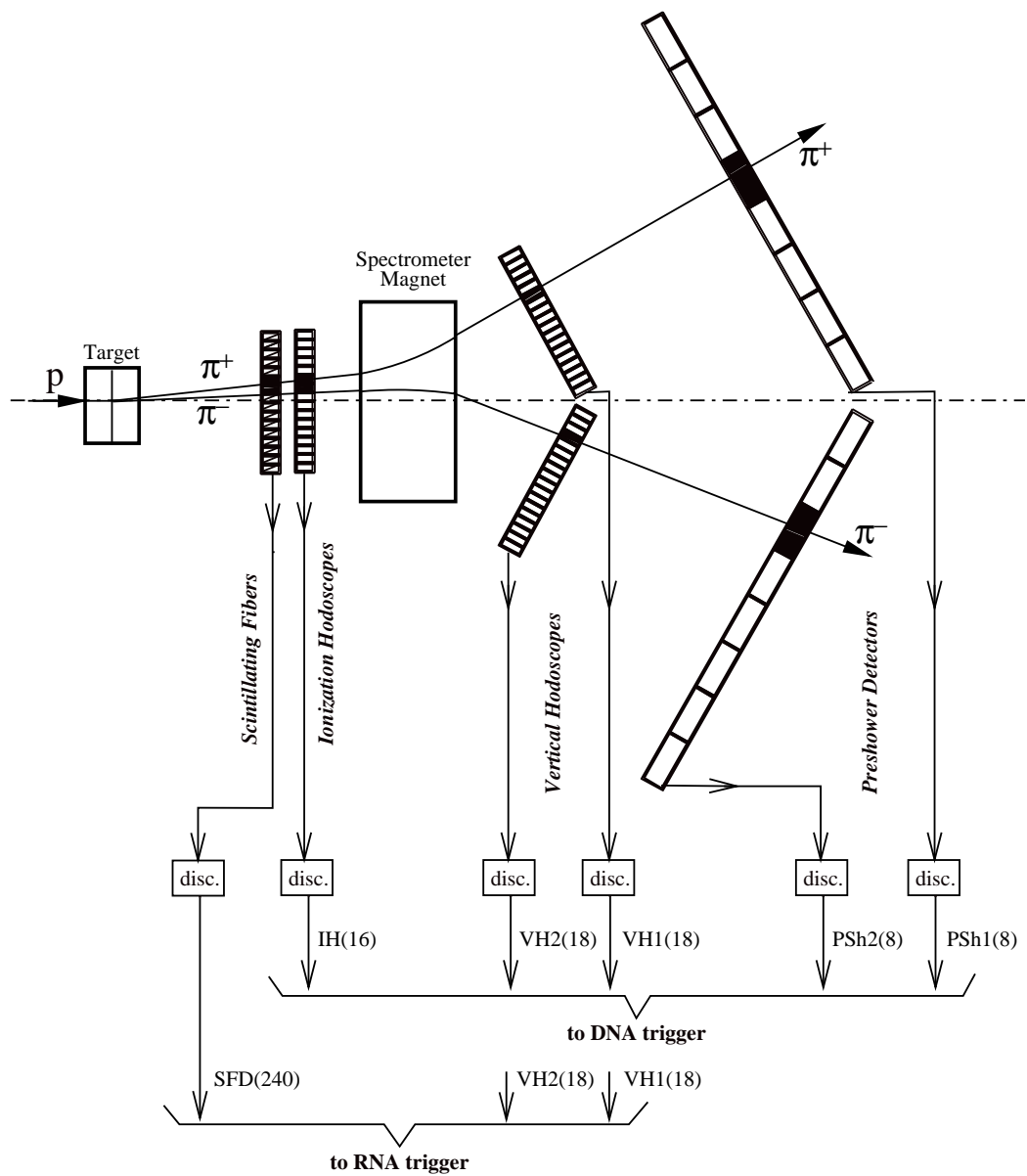,width=\textwidth}
\end{center}
\caption{DIRAC detectors used for the neural network triggers DNA and
  RNA. Numbers of signals from each detector are given in parentheses.}
\label{rna}
\end{figure}

DNA receives the hit patterns from the vertical hodoscopes VH1, VH2,
the X-planes of the ionization hodoscope IH and optionally the
preshower detectors PSh1 and PSh2. In contrast to T3, DNA uses a
straight-forward IH hit-map (i.e. without double ionization selection
by IH logic).  The detectors used for that trigger are shown in
Fig.~\ref{rna}.

DNA is able to handle the events with up to 2 hits in each
vertical hodoscope VH and up to 5 hits in each IH X-plane.
If the number of hits exceeds these values in any of these
detectors, DNA accepts the event for further off-line evaluation.
In case there is only 1 hit in an IH plane, it is assumed that two
particles cross the same IH slab.

Each IH plane is used independently in combination with both VHs.
Therefore DNA consists of two identical parts. Each DNA part is based
on three electronic modules: the interface and decision card
\cite{IDC}, the neural network cards \cite{NN} and the POWER-PC VME
master CPU card (Motorola MVME\,2302).  The subdecisions of the two
parts are combined in a logical OR to minimize inefficiency due to
gaps between the IH slabs.

The neural network was trained to select particle pairs with low
relative momenta: $Q_x < 3\,$MeV/$c$, $Q_y < 10\,$MeV/$c$ and $Q_L <
30\,$MeV/$c$. The events which do not satisfy any of these conditions
are considered ``bad'' and rejected. The training of the system was
done initially with Monte-Carlo simulated events and then, before the
implementation in the DIRAC trigger system, checked with real
experimental data.

The DNA logic starts with T0 and in about 210~ns evaluates an event.
Since DNA does not take into account any Cherenkov or horizontal
hodoscopes information, an event selected by DNA is only further
processed if it is also accepted by the T1 (or T1$\cdot$T2) trigger.
The DNA rejection is approximately 2.3 in respect to T1. Its
efficiency in the low $Q$ region is 94\%.

To increase the selection efficiency, the DNA logic at the later stage
of the experiment was supplemented with the RNA trigger system. The
RNA operation is similar to that of DNA. Instead of the IH data, RNA
uses the information from the X-plane of the scintillating fiber
detector SFD. Figure~\ref{rna} shows the related detectors used.  Finer
granularity upstream the magnet (0.5 mm in SFD in comparison with 6 mm
in IH) provides higher trigger efficiency for pion pairs with small
opening angles. The RNA decision time is 250~ns.

The OR between DNA and RNA provides a rate rejection of 1.9--2.0, at
the same time increasing the efficiency in the low momentum $Q$ range
to 99\%.

\section{Trigger 4}

Trigger 4 is the final trigger stage. It reconstructs straight tracks
in the X-projection of the drift chambers and analyses them with
respect to relative momentum. T4 starts with a T1 trigger signal.

\begin{figure}[htbp]
\begin{center}
\epsfig{file=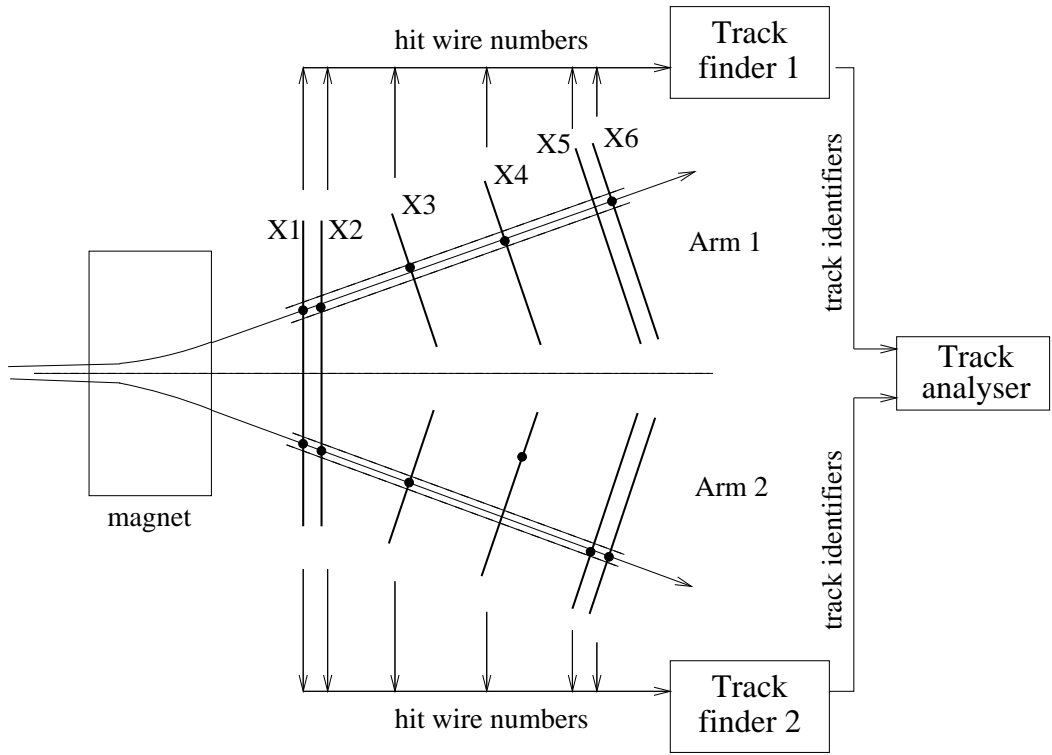,width=\textwidth}
\end{center}
\caption{T4 operation block diagram. Only the drift chamber
X-planes involved in T4 are shown.}
\label{t4logic}
\end{figure}

T4 includes two stages: the track finder and the track analyser. The
track finder (an identical processor is used for each arm) receives
the numbers of the fired wires from all drift chamber X-planes. Drift
time values are not used in the T4 logic.  The block diagram of the T4
operation is shown in Fig.~\ref{t4logic}.

The track finder operation is based on an endpoint algorithm.  Drift
chamber planes X1 (or X2) and X5 (or X6) are the base planes for the
track search (in total there are 6 X-planes in each spectrometer arm).
Every pair of the hits in the base planes is considered as end points
of a straight line. Each of allowed hit combination is used to select
a pattern window for the intermediate planes. A track is assumed to be
found if the number of hits within the window exceeds a preset value.
The window width and position can be set for every plane
independently. The minimum number of hits per track is also variable
(a typically used value is 4).  A unique number, ``track identifier'',
which contains the encoded numbers of the fired wires in the base
planes, is ascribed to the found track.  Parasitic combinations
(i.e. repeated track identifiers) are suppressed by the track finder.

If tracks are found in both arms, the track analyser continues the
event evaluation. The look-up memory of the track analyser contains
all possible combinations of the track identifiers for pion pairs with
$Q_L \leq30\,$MeV/$c$ and $Q_x \leq3\,$MeV/$c$. These ``allowed''
combinations are obtained with simulation using the precise geometry
of the setup. The track analyser receives the track identifiers from
both arms and compares them with the content of the look-up memory. If
a relevant combination is found, the T4 processor generates a positive
decision signal which starts the data transfer to the VME buffer
memories. Otherwise, the Clear and Reset signals are applied to the
DAQ and trigger systems.

T4 is able to operate in different modes which are defined
by several parameters loaded at the start of the data taking run.
In the standard mode:
\begin{itemize}
\item Presence of the positive decision of the DNA/RNA trigger at the
  T4 input is mandatory;
\item A special ``transparent'' input is activated. If the signal at
  this input is present when T4 starts, T4 always generates a positive
  decision. The calibration trigger marks are sent to this input thus
  disabling their rejection;
\item Timeout for the T4 decision is enabled. If a decision is not
  reached before the end of the timeout interval, a positive decision
  is generated unconditionally.
\end{itemize}

The T4 decision time is not fixed and depends on the complexity of the
event. Being in average around 3.5~$\mu$s, it varies from less than
1.5~$\mu$s for simple events to more than 20~$\mu$s for the most
complicated ones. To avoid large dead time losses (and due to some
restrictions imposed by MSGC Clear process) the timeout interval is
set to 10 $\mu$s.

The track finder stage provides additional rejection of 10--20\%
events depending on background conditions. Rejected events are ones
without tracks in DC X-planes. The main part of the rate reduction
comes from the track analyser. The complete T4 provides a rejection
factor of around 5 with respect to the T1 rate or around 2.5 with
respect to DNA/RNA.  The T4 efficiency in the low $Q$ region exceeds
99\%.

\section{Trigger control and operation}

The whole trigger system is fully computer controlled: no manual
operations are needed in order to modify the trigger conditions.  The
developed trigger software allows one to vary the electronic logic or
the front-end settings. At the beginning of the data taking phase the
corresponding parameters of all electronic modules are loaded using
the files which define the status of the front-end and trigger
electronics in accordance with the selected trigger (thresholds,
enabled trigger levels or modes, prescaling factors for different
calibration triggers etc.).

The complete trigger system operates in the following way.  In the
first trigger version, Fig.~\ref{trigab}a, a coincidence of the T1 and
T2 triggers starts digitization in the FERA \cite{FERA} modules (ADC,
TDC etc.) and in the drift chamber electronics. The positive T3 and
DNA decisions joined in a logical OR trigger the MSGC electronics and
allow the readout of the whole event to buffer memories.  When an
event is rejected, the data in the FERA and DC branches are cleared
and no readout takes place; no Clear is applied to MSGC as they were
not started yet. The T1$\cdot$T2 signal inhibits generation of a new
T1 and T0 until the end of the Clear processes followed by either
readout or discarding of the data after a negative decision.

At the typical experimental conditions quoted in Sec.~\ref{sec:intr}
the T1 rate is close to 5000/spill, the rate of accepted events is
1800--2000/spill (the ratio of the rates exceeds the reduction factor
1.9 of T3+DNA due to the dead time of the data acquisition and trigger
systems). The calibration $e^+e^-$, $\Lambda$ and $K$ triggers (used
with the prescaling factors 7, 3 and 2, respectively) constitute 15\%
of the accepted event sample.  With these rates an overflow of the
buffer memories in some readout branches occurs occasionally and the
rate capability of the data acquisition system is near the limit when
the setup receives several beam bunches close in time.  To eliminate
these drawbacks a different trigger version, Fig.~\ref{trigab}b, is
used.

In this revised trigger version the drift chamber processor T4 and the
additional neural trigger chain RNA were implemented.  The T2 and T3
stages were skipped as they did not add considerably to the rate or
dead time reduction anymore.

In Fig.~\ref{trigab}b digitization starts with T1.  Similarly to the
previous scheme, in case of the negative decision of DNA+RNA the data
in the FERA and $DC$ branches are cleared.  The positive DNA+RNA
decision starts the MSGC electronics and gives the confirmation to T4
to begin processing.
        
The negative T4 decision leads to Clear of data in all branches
including MSGC. Note that MSGC clearing takes more time than that in
any other branch and for this reason MSGCs start later. Therefore dead
time losses from the MSGC clear process occur only when T4 rejects an
event already accepted by DNA+RNA.  All events accepted by T4 are
transferred to the VME memories to be read out for further off-line
analysis.

At the same T1 rate of 5000/spill the DNA+RNA rate is 2000--2200/spill
and the T4 rate of accepted events is around 700/spill. The sum of the
calibration triggers is around 20\% at the applied prescaling factors
14, 6 and 4 for $e^+e^-$, $\Lambda$ and $K$-triggers, respectively.

\begin{figure}[htbp]
\begin{center}
\epsfig{file=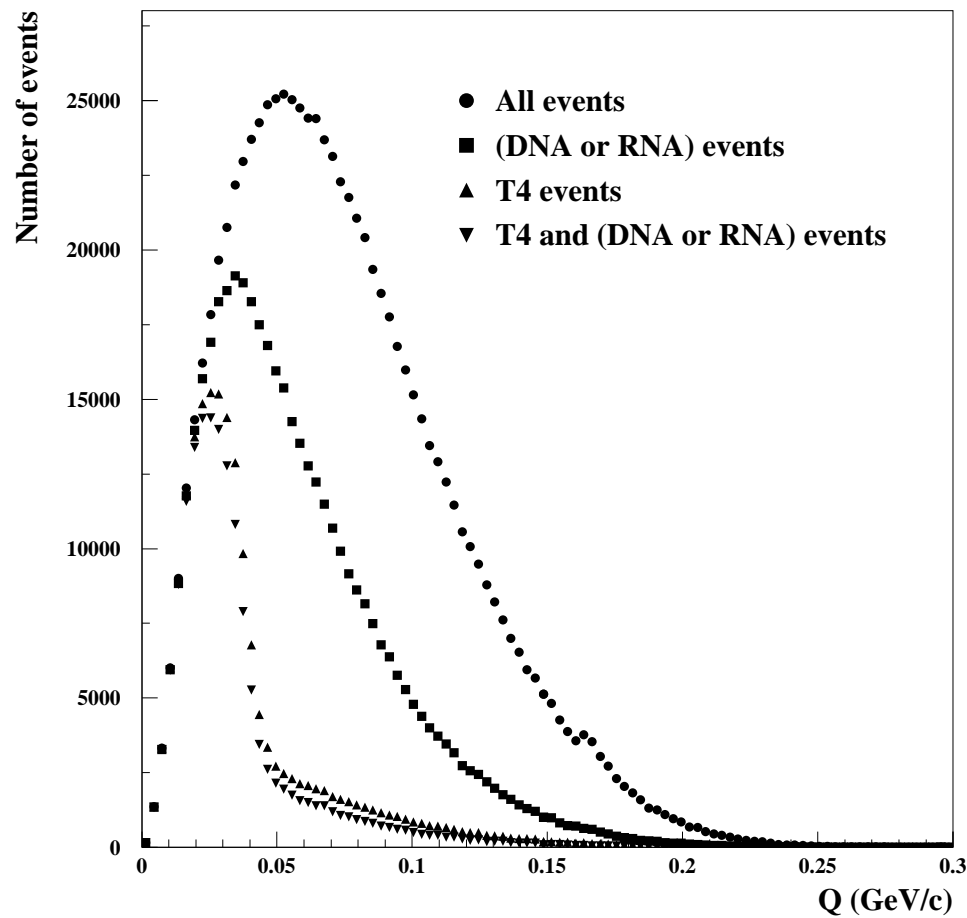,width=\textwidth}
\end{center}
\caption{Total relative momentum $Q$
for events selected by different trigger levels.}
\label{Qtot}
\end{figure}

\begin{figure}[htbp]
\begin{center}
\epsfig{file=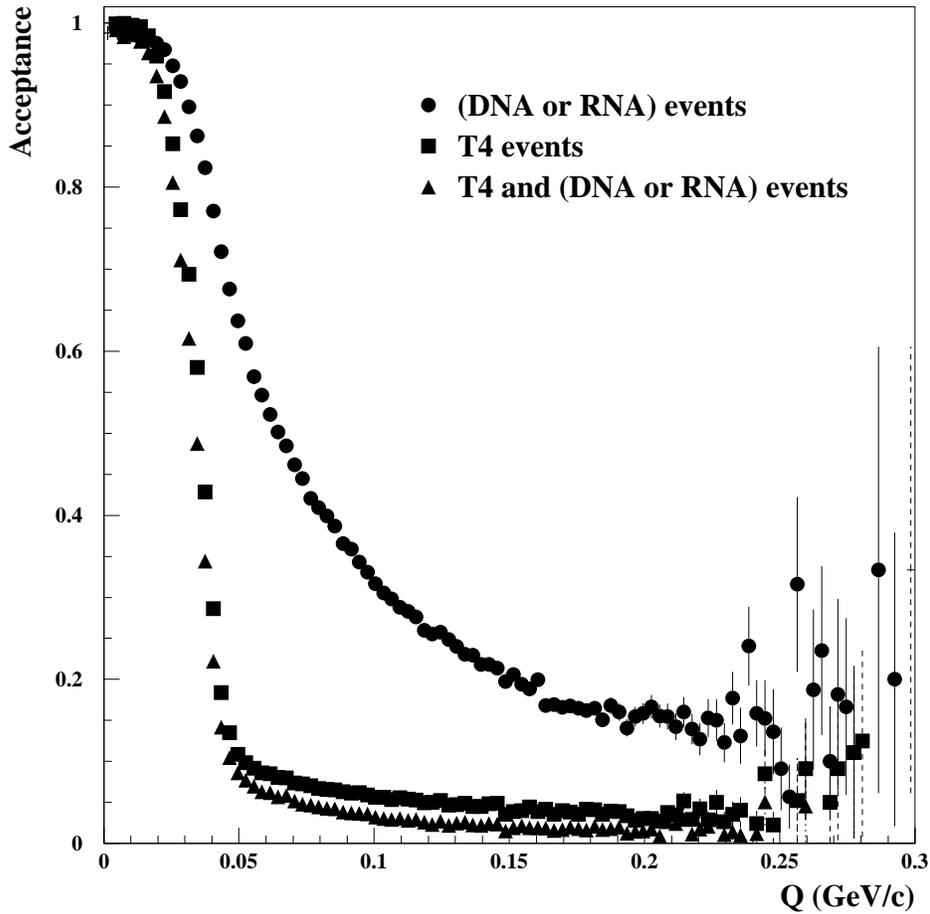,width=\textwidth}
\end{center}
\caption{Efficiency of DNA+RNA, T4 and (DNA+RNA)$\cdot$T4 triggers
  with respect to T1 as a function of total $Q$.}
\label{Qteff}
\end{figure}

\begin{figure}[htbp]
\begin{center}
\epsfig{file=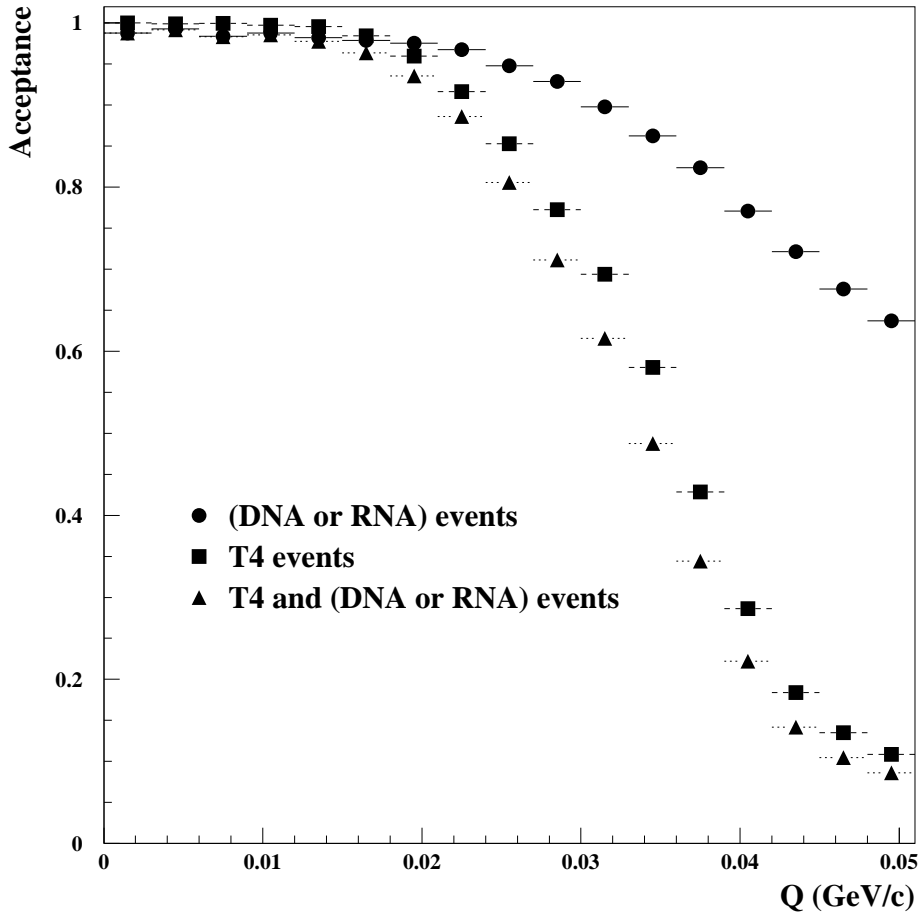,width=\textwidth}
\end{center}
\caption{Same as in Fig.~\ref{Qteff} with an expanded $Q$ scale.}
\label{Qeffexp}
\end{figure}

\begin{figure}[htbp]
\begin{center}
\epsfig{file=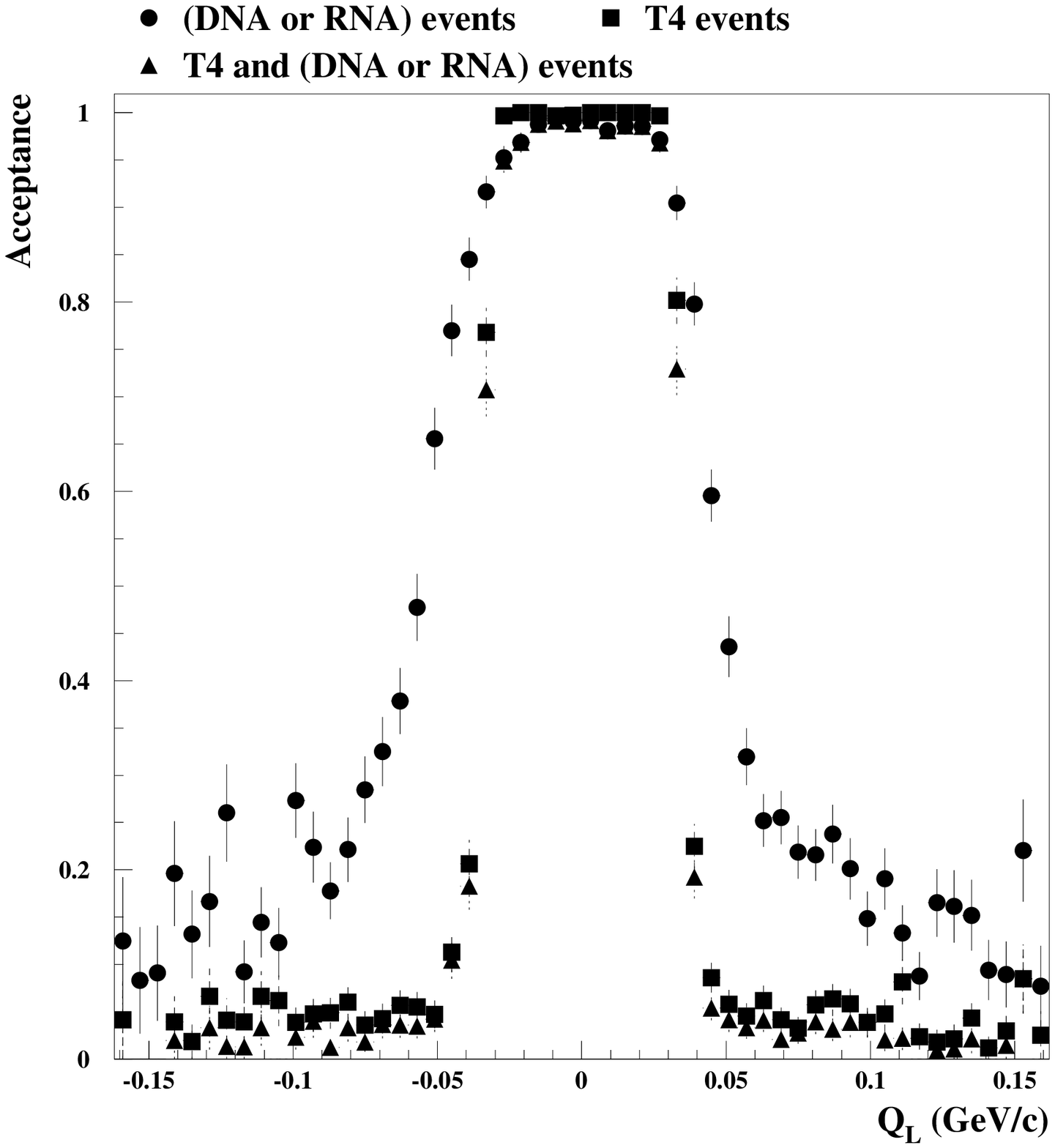,width=\textwidth}
\end{center}
\caption{Efficiency of DNA+RNA, T4 and (DNA+RNA)$\cdot$T4 triggers
  with respect to T1 as a function of $Q_L$ for events with
  $Q_x<3\,$MeV/$c$ and $Q_y<3\,$MeV/$c$.}
\label{Qleff}
\end{figure}

The performance of the trigger is illustrated in
Figs.~\ref{Qtot}--\ref{Qleff}.  In Fig.~\ref{Qtot} the acceptance of
different triggers as a function of the relative momentum $Q$ is
presented. The curves for events selected by T1, T1$\cdot$(DNA+RNA), T1$\cdot$T4
and for the complete trigger T1$\cdot$(DNA+RNA)$\cdot$T4 are shown. In
Figs.~\ref{Qteff} and \ref{Qeffexp} the efficiencies of triggers are
shown with respect to T1 (the ratios of the curves in Fig.~\ref{Qtot}
to the T1 curve). The efficiency for selecting low $Q$ events by the
trigger processors is evident.

The efficiencies of triggers as a function of the longitudinal
component of the relative momentum, $Q_L$, when $Q_x < 3\,$MeV/$c$ and
$Q_y< 3\,$MeV/$c$ are presented in Fig.9. The T4 efficiency alone with
respect to T1 is over 99\% in the whole region $|Q_L| < 30\,$MeV/$c$.
Common selection by DNA+RNA and T4 results in the 98\% efficiency for
$|Q_L| < 22\,$MeV/$c$ and 95\% for $|Q_L| < 30\,$MeV/$c$.

The complete trigger provides a factor of 1000 in rate reduction with
respect to the single counting rates of the downstream detectors
keeping a high efficiency in the range of low pair relative
momenta.

\section{Trigger quality control}

Improper trigger functioning could lead to losses of events or
systematic biases in the collected data which cannot be corrected
later at the off-line analysis. For that reason the trigger quality is
under permanent control.

The counting rates of different triggers and their submodes at all
trigger levels are measured by CAMAC scalers. Their content is read
out and recorded at the end of every accelerator spill and is also
displayed by the on-line monitoring program. This provides control
over the trigger stability.

The trigger efficiency is under control as well. One of the TDC
modules (the ``trigger register'') is intended for detection of trigger
marks.  Each decision of every trigger level, its submode type and
arrival time per event are recorded. This information is available for
the off-line analysis and on-line monitoring.

To test the efficiency of all higher than T1 triggers, their rejection
power and acceptance in terms of the relative momentum $Q$, data are
periodically collected with T1 as the only active trigger.  The other
triggers perform their analysis and send their trigger marks to the
trigger register (T4 decisions, due to its longer processing time, are
written in a special ``mail box'' in the DC buffer memory). A regular
off-line express-analysis permits the detection of any deviations in
the trigger performance. The results presented in
Figs.~\ref{Qtot}--\ref{Qleff} were obtained just from such data
collected with T1 trigger.

To check the operation of the whole trigger system, including the
primary T1 and T0 steps, a special minimum bias trigger is sometimes
used. Two minimum bias trigger options are available with the simplest
logic: IH$\cdot$VH1 or IH$\cdot$VH2, i.e. with only one upstream and one
downstream detector involved. The events thus triggered may contain
hits in other detectors as well and eventually even other trigger
processors may select them. In these cases all the corresponding
trigger marks should be generated.  Off-line analysis of the data from
all the detectors together with the recorded trigger marks makes it
possible to test the trigger efficiency at all trigger stages.

One more trigger is arranged for tests of the front-end, trigger and
readout electronics. Artificially generated signals are sent to inputs
of the front-end electronics of all the detectors (except SFD, MSGC
and DC). Relative delays between the generator signals coming to
different detector groups are adjusted to be like those from real
particles. As a result, all electronic channels in the trigger and
readout modules can be tested.

\section{Conclusions}

The trigger system of DIRAC has been operating since 1999.  It passed
through several upgrade steps and at present provides considerable
trigger rate reduction which satisfies the rate capabilities of the data
acquisition system and the volume of the buffer memories used.  The
system proved to be reliable and convenient for the users.

\section*{Acknowledgments}
The authors are grateful to L.Nemenov for coordination of trigger
activity, V.Olshevsky and S.Trusov for readout software support,
J.Buytaert and V.Korolev for development of some electronic units and
to many members of the DIRAC collaboration for participation in
trigger tuning and control during the beam measurements.  The work was
supported by the Swiss National Science Foundation and by the Russian
Foundation for Basic Research, project 01--02--17756.


\begin{thebibliography}{99}
\bibitem{proposal} B.Adeva et al.:
\emph{Lifetime measurement of $\pi^+ \pi^-$ atoms to test low energy
QCD predictions},
Proposal to the SPSLC, CERN/SPSLC 95-1,
SPSLC/P 284, December 1994.

\bibitem{Trig1} L.Afanasyev, M.Gallas, V.Karpukhin, A.Kulikov,
NIM, A479 (2002) 407--411.

\bibitem{trig2} F.Takeutchi: \emph{Study of the trigger 
logic with front-end detectors II}, Report at the DIRAC 
trigger/electronics meeting, CERN, February 1996.

\bibitem{SFDtrig} V.Agoritsas et al.,
NIM A411 (1998) 17--30.

\bibitem{peaksens} A.Gorin et al., 
NIM A452 (2000) 280--288.

\bibitem{T3note} V.Karpukhin, A.Kulikov, V.Yazkov: \emph{Third 
level trigger for DIRAC. Versions of implementation},
DIRAC Note 96-27, CERN, 1996.

\bibitem{Trig3} M.Gallas: \emph{The Complete Software-Programmable 
Third Level Trigger for DIRAC}, 
Application Note AN-60, LeCroy Corporation, 1999.\\
M.Gallas: \emph{Third level trigger of the DIRAC experiment},
To be published in NIM A482/1-2.

\bibitem{DNA} P.Kokkas, M.Steinacher, L.Tauscher, S.Vlachos,
NIM A471 (2001) 358--367.

\bibitem{neural} F.R.Leimgruber et al., 
NIM A365 (1995) 198--202.

\bibitem{IDC} M.Steinacher: \emph{Interface and decision card;
the design, Revision 2.0}, Internal Report, University of Basel,
1999.

\bibitem{NN} M.Steinacher: \emph{Hardware implementation of a fast
neural network}, In Proceedings of the Sixth International
Workshop on Software Engineering, Artificial Intelligence and
Expert Systems (AIENP 1999), Parisianou Press, Heraclion, 2000.

\bibitem{FERA}
\emph{Fast Encoding and Readout ADC System Possibilities},
Application Note AN-4004A, LeCroy Corporation.
\end{thebibliography}
\end{document}